\documentclass[aps,pre,twocolumn,showpacs,superscriptaddress,groupedaddress]{revtex4}  
\usepackage{amsfonts}
\usepackage{amsmath}
\usepackage{amssymb}
\usepackage{version}
\usepackage{graphicx}%
\includeversion{New_connection}
\excludeversion{Old_connection}

\begin{document}
\title[Verlet Revision]{A simple and effective Verlet-type algorithm for simulating Langevin dynamics}
\author{Niels Gr{\o}nbech-Jensen}
\affiliation{Department of Mechanical and Aerospace Engineering, University of California, Davis, CA 95616}
\affiliation{Department of Chemical Engineering and Materials Science, University of California, Davis, CA 95616}
\affiliation{Computational Research Division, Lawrence Berkeley National Laboratory, Berkeley, CA 94720}
\author{Oded Farago}
\affiliation{Department of Mechanical and Aerospace Engineering, University of California, Davis, CA 95616}
\affiliation{Department of Biomedical Engineering, Ben Gurion University of the Negev, Be'er Sheva, 84105 Israel}
\affiliation{Ilse Katz Institute for Nanoscale Science and Technology, Ben Gurion University of the Negev, Be'er Sheva, 84105 Israel}

\begin{abstract}
We present a revision to the well known St{\"o}rmer-Verlet algorithm for
simulating second order differential equations. The revision addresses
the inclusion of linear friction with associated stochastic noise, and
we analytically demonstrate that the new algorithm correctly
reproduces diffusive behavior of a particle in a flat potential. For a harmonic
oscillator, our algorithm provides the exact Boltzmann distribution for any
value of damping, frequency, and time step for both underdamped and overdamped
behavior within the usual stability limit of the Verlet algorithm.
Given the structure and simplicity of the method we conclude that this
approach can trivially be adapted for contemporary applications, including
molecular dynamics with extensions such as molecular constraints.
\end{abstract}
\maketitle

\section{Introduction}

In Molecular Dynamics (MD) simulations, Newton's equations of motion
are solved numerically to produce trajectories of the system in the
microcanonical $(N,V,E)$ ensemble. The most commonly used scheme for
this purpose is the Verlet method, which is based on truncated Taylor
expansions for the evolution of a particle with mass $m$, coordinate $r(t)$, velocity $v(t)$, and force $f(r,t)$
\cite{Frenkel_1996,Swope_1982}. Introducing the discrete-time variables $r^n=r(t_n)$,
$v^n=v(t_n)$, and $f^n=f(r^n,t_n)$, the so-called velocity explicit Verlet
(or velocity Verlet) scheme reads
\begin{eqnarray} 
r^{n+1}&=&r^n+v^ndt+\frac{dt^2}{2m}f^n \label{vverlet_pos}\\
v^{n+1}&=&v^n+\frac{dt}{2m}(f^n+f^{n+1}) \label{vverlet_vel}.
\end{eqnarray}
By
considering two successive time steps and eliminating the velocity
variables from the equations, the more original form of the St{\"o}rmer-Verlet
method \cite{Stormer_1921,Verlet_1967} (here called the position-Verlet method), is found
\begin{eqnarray} 
r^{n+1}=2r^n-r^{n-1}+\frac{dt^2}{m}f^n,
\end{eqnarray}
with the associated velocity calculated by the central difference
\begin{eqnarray}
v^n & = & \frac{r^{n+1}-r^{n-1}}{2dt}.
\end{eqnarray}
The Verlet scheme is accurate to second order in $dt$, which means
that the deviation, per time step, of the computed trajectory from the
true (analytic) one, scales with the third power of $dt$. The merits of the
Verlet scheme that make it so widely popular for MD simulations
include its convenience, efficiency, and time reversibility, which ensures
that the error of the total energy of long-time integrations is bounded and
does not drift or diffuse, as illustrated by the exact solution to the Verlet method
applied to a harmonic oscillator (see, e.g., below).

The most frequently used ensemble in statistical-mechanics is the
canonical $(N,V,T)$ ensemble where the temperature of the system,
rather than its energy, is constant. A variety of methods for conducting MD
simulations in the canonical ensemble have been proposed over the
years. A very appealing class of such methods include integrators for
Langevin Dynamics (LD) simulations. In LD, two forces are
added to the conservative force field - a friction force proportional
to the velocity with friction coefficient $\alpha\ge0$, and thermal
white noise $\beta(t)$. Explicitly, the Langevin equation of motion is
given by \cite{Parisi_1988}
\begin{eqnarray}
\dot{r} & = & v   \label{Eq_L2}\\
m\dot{v} & = & f(r,t)-\alpha v+\beta(t) \, . \label{Eq_L1}
\end{eqnarray}
In order to satisfy the dissipation-fluctuation theorem, it is often assumed
that the stochastic force is Gaussian distributed, and has the
statistical properties \cite{Parisi_1988}
\begin{eqnarray}
\langle\beta(t)\rangle & = & 0 \label{Eq_L3}\\
\langle\beta(t)\beta(t^\prime)\rangle & = & 2\alpha k_BT\delta(t-t^\prime),  \label{Eq_L4}
\end{eqnarray}
where $k_B$ is Boltzmann's constant and $T$ is the thermodynamic
temperature of the heat bath.

The difficulty to develop accurate integrators for LD stems from the
non-analytic nature of $\beta(t)$, which invalidates the Taylor
expansion commonly used for the derivation of the Verlet
scheme. The most naive way to overcome this difficulty is to replace
the delta-function correlated noise with a set of rectangular pulses
of mean-squared size $\sqrt{2k_BT\alpha/dt}$, each of which acting
over the centered time interval $(t_n-dt/2, t_n+dt/2)$. Employing this
approximation for $\beta(t)$ yields the famous
Brooks-Br\"{u}nger-Karplus (BBK) scheme \cite{Brunger_1984}, which unfortunately
turns out to exhibit a simulated temperature that differs
by ${\cal O}(dt)$ from the correct one \cite{Skeel_2003}. This disappointing result
can be attributed to the effect of a small time step $dt$, which results in a 
stochastic velocity change of order $\sqrt{dt}$ that overshadows the ``regular'' (deterministic)
linear velocity change of order $dt$. Statistical analysis of the small time interval shows, in fact, that 
the opposite is true, since the average over all noise
realizations vanishes. In order to develop a reliable integrator for LD one
therefore needs to carefully treat the coupling between the stochastic and
analytic contributions. This is, unfortunately, not easily accomplished
with a discretized approximation for $\beta(t)$. A different approach
to the problem has been introduced by Ermak and Buckholz (EB)
\cite{eb_1980}. The EB method is based on a numerical integration of the
formal solution to the Langevin equation. This gives
\begin{widetext}
\begin{eqnarray}
v^{n+1}&=&v^n\exp(-\alpha
dt/m)+\frac{1}{m}\int_{t_n}^{t_{n+1}}\exp\left[-\alpha
(t_{n+1}-t^\prime)/m\right]\left[f(t^\prime)+\beta(t^\prime)\right]\,dt^\prime\label{analytic_v}\\
r^{n+1}&=&r^n+\frac{mv^n}{\alpha}\left[1-\exp\left(-\alpha dt/m\right)\right]+\frac{1}{\alpha}\int_{t_n}^{t_{n+1}}\left\{1-\exp\left[-\alpha(t_{n+1}-t^\prime)/m\right]\right\}\left[f(t^\prime)+\beta(t^\prime)\right]\,dt^\prime.\label{analytic_r}
\end{eqnarray}
\end{widetext}
Several features distinguish the EB method from Verlet-type
schemes such as BBK. First, the friction force is accounted for
separately from the other two forces, via exponentially decaying
functions. Second, this scheme considers an ensemble of trajectories
instead of a single one, thereby requiring two random Gaussian
numbers per time step. The one appearing in the velocity equation,
$\int_0^{dt} \exp[-\alpha(dt-t^\prime)/m]\beta(t^\prime)dt^\prime$, represents the
stochastic change in velocity during the time interval. The other,
$\int_0^{dt}\{1-\exp[-\alpha(dt-t^\prime)/m]\}\beta(t^\prime)dt^\prime$, appears in the
position equation and characterizes the random displacement. These two
numbers are different, but {\it correlated}.
Third, there
is formally no need to compute the new position before the new
velocity as in the velocity-Verlet scheme and, in principle, one can
interchange the order by which the equations are evolved. The last
point is particularly important since it is directly related to the most critical
issue that remains unsolved with the EB approach, which is how to
perform the numerical integration over the deterministic force
$f(t)$. The appearance of exponential ``weight functions'' in the
integrals in Eqs.~(\ref{analytic_v}) and (\ref{analytic_r}) opens the possibility
for using a variety of linear combinations involving $f^{n-1}$,
$f^n$, and $f^{n+1}$. The most popular integration schemes for the
deterministic force constitute the van Gunsteren-Berendsen \cite{vgb_1982}
and the Langevin impulse \cite{Skeel_2002} methods.

More recently, Ricci and Ciccotti (RC) introduced a formalism for a
systematic derivation of numerical integrators for LD \cite{rc_2003}.
The essential steps of the formalism are to (i) express the integrator as
an exponential operator, (ii) split the time interval $dt$ into a
series of even smaller time steps, and (iii) use the Suzuki-Trotter
expansion for time-ordered exponential operators to find an
approximation for the evolution operator corresponding to $dt$.  By
using mid-point splitting, RC arrived at a scheme that require only
one random number per $dt$. The same formalism was later used
to develop a new family of integrators that more carefully treat the
coupling between the stochastic and deterministic components of the
dynamics \cite{Melchionna_2007,Bussi_2007}. A characteristic feature of these new
schemes is that they require two {\em independent} random numbers
per time step, which is different from the EB family of schemes, where
two correlated random numbers are involved.

Returning for a moment to MD simulations of Newtonian Dynamics, it is
important to recall that, due to its computational efficiency and time
reversibility, the Verlet method is regarded as superior to other
integration methods that may produce more accurate
trajectories. In the case of LD, the trajectories include a random
component and, therefore, their accuracy (which can only be
defined in statistical terms) becomes an issue of even less
importance. The efficiency of different methods for canonical ensemble
simulations of many-particles systems must be tested according to
their ability to reproduce measurable statistical quantities, such as the
Boltzmann distribution corresponding to
the temperature of the heat bath. To this end, none of the existing
methods has managed to demonstrate exact thermodynamic response
\cite{Skeel_2003}. We here present a novel Verlet-type
scheme for LD simulations that is very simple to implement and which
yields {\it correct} statistical-mechanical behavior of a particle
diffusing in both a flat and a harmonic potential.

\section{Derivation of the algorithm}

In the spirit of the simplicity of the Verlet algorithm in its
traditional forms, we here arrive at a new useful method through a
straight-forward derivation.  Starting with the continuous-time
Langevin equations (\ref{Eq_L2})-(\ref{Eq_L4}), we integrate
Eq.~(\ref{Eq_L1}) over a (small) time interval $dt$ between two times,
$t_n$ and $t_{n+1}=t_n+dt$:
\begin{eqnarray}
\int_{t_n}^{t_{n+1}}m\dot{v} \, dt^\prime& = & \int_{t_n}^{t_{n+1}}f \, dt^\prime -\int_{t_n}^{t_{n+1}}\alpha\dot{r} \, dt^\prime \nonumber \\ && + \int_{t_n}^{t_{n+1}}\beta(t^\prime) \, dt^\prime \;
. \label{Eq_intL1}
\end{eqnarray}
With no approximation, this can be written
\begin{eqnarray}
m(v^{n+1}-v^n) & = & \int_{t_n}^{t_{n+1}}f \, dt^\prime-\alpha(r^{n+1}-r^n)
+ \beta^{n+1},
\label{Eq:init_rVv_prel} 
\nonumber \\
\end{eqnarray}
where 
\begin{eqnarray}
\beta^{n+1} & \equiv & \int_{t_n}^{t_{n+1}}\beta(t^\prime) \, dt^\prime  
\end{eqnarray}
is a Gaussian random number with $\langle\beta^n\rangle=0$ and
$\langle\beta^n\beta^l\rangle=2\alpha k_BTdt\delta_{n,l}$.

Integrating over Eq.~(\ref{Eq_L2}) yields
\begin{eqnarray}
 \int_{t_n}^{t_{n+1}}\dot{r} \, dt^\prime & = &  r^{n+1}-r^n \; = \; \int_{t_n}^{t_{n+1}}v \, dt^\prime . \label{Eq_int_av}
\end{eqnarray}
which can be approximated with
\begin{eqnarray}
r^{n+1}-r^n & \approx & \frac{dt}{2}(v^{n+1}+v^n) \;
. \label{Eq:init_rVp_prel}
\end{eqnarray}
An equation similar to (\ref{Eq:init_rVp_prel}) has been used by Ricci
and Ciccotti in one of the forms of their scheme (see Eq.(18) in
Ref.~\cite{rc_2003}). It introduces errors that scale with $dt^3$ in
the deterministic trajectory and in the variance of the stochastic
component (see discussion in Ref.~\cite{thalmann_2007}).

Inserting $v^{n+1}$ from Eq.~(\ref{Eq:init_rVv_prel}) into
Eq.~(\ref{Eq:init_rVp_prel}) provides a convenient pair of equations
\begin{eqnarray}
r^{n+1}-r^n & = & b dt \, v^n+\frac{bdt}{2m}\int_{t_n}^{t_{n+1}}f \, dt^\prime +\frac{bdt}{2m}\beta^{n+1} \label{rVp_almost}\\
v^{n+1}-v^n & = & \frac{1}{m}\int_{t_n}^{t_{n+1}}f \, dt^\prime -\frac{\alpha}{m}(r^{n+1}-r^n) + \frac{1}{m}\beta^{n+1}  \label{rVv_almost}
\nonumber \\
\end{eqnarray}
where 
\begin{eqnarray}
b\equiv\frac{1}{1+\frac{\alpha dt}{2m}}.
\label{b_definition}
\end{eqnarray}
For any given realization of $\beta^n$, Eq.~(\ref{rVp_almost}) is
correct to second order in $dt$, while (\ref{rVv_almost}) is exact as
written. 
We now approximate the integral of the
deterministic force $f$ such that both equations are correct to second
order in $dt$:
\begin{widetext}
\begin{eqnarray}
r^{n+1}&=&r^n + b dt \, v^n+\frac{bdt^2}{2m}f^n +\frac{bdt}{2m}\beta^{n+1} \label{rVp_vs1}\\
v^{n+1}&=&v^n + \frac{dt}{2m}(f^n+f^{n+1})-\frac{\alpha}{m}(r^{n+1}-r^n) + \frac{1}{m}\beta^{n+1} . \label{rVv_vs1}
\end{eqnarray}
\end{widetext}
We first notice that when $\alpha=0$, the above equations
(\ref{rVp_vs1})-(\ref{rVv_vs1}) reduce to the standard velocity-Verlet
scheme given in Eqs.~(\ref{vverlet_pos}) and (\ref{vverlet_vel}). Second, we notice
the very reasonable feature of the representation of the damping,
which is calculated as the integral of the actual path that the object
has traveled during the time step $dt$. Third, the noise is
represented as a {\it single} stochastic variable for each time step,
realized by a single aggregated impulse during $dt$ that influences
the dynamics over the time step.  In this regard,
Eqs.~(\ref{rVp_vs1}) and (\ref{rVv_vs1}) constitute a simple functional
Verlet-type scheme for solving stochastic Langevin equations. Unlike
the aforementioned EB-type family of schemes
\cite{eb_1980,vgb_1982,Skeel_2002}, our method does not employ two
stochastic variables. This is a consequence of our
main objective; namely to produce the correct statistical-mechanics for
large ensembles (spatial or temporal), while focusing less on the
detailed dynamics  {\em within} each time step.

Before proceeding to analyzing the behavior of the developed scheme, we
rewrite the method in a couple of equivalent different forms that
can be useful and that can illustrate close connections to previously published
work. First, we observe that by inserting
Eq.~(\ref{rVp_vs1}) into (\ref{rVv_vs1}), the equations can be rewritten
\begin{eqnarray}
r^{n+1} & = &
r^n+bdtv^n+\frac{bdt^2}{2m}f^n+\frac{bdt}{2m}\beta^{n+1}\label{rVp_vs2}\\
v^{n+1} & = &
av^n+\frac{dt}{2m}\left(af^n+f^{n+1}\right)+\frac{b}{m}\beta^{n+1}
\label{rVv_vs2}
\end{eqnarray}
where 
\begin{eqnarray}
a&\equiv&\frac{1-\frac{\alpha dt}{2m}}{1+\frac{\alpha dt}{2m}}.
\label{a_definition}
\end{eqnarray}
This form reveals that the derived method presented here parallels one
mentioned in passing by Melchionna (Eq.~(41) in
Ref.~\cite{Melchionna_2007}), where the contribution of the friction
force has been integrated similarly, but with a different noise term
that includes two independent random numbers per time step. Second, we
rewrite the scheme in the form given only by displacement
coordinates. This is accomplished by subsequently inserting the
expressions for $v^n$ (\ref{rVv_vs2}) and $r^n$
(\ref{rVp_vs2}) into that of $r^{n+1}$ given in Eq.~(\ref{rVp_vs2}).
The result is:
\begin{eqnarray}
r^{n+1} & = &
2br^n-ar^{n-1}+\frac{bdt^2}{m}f^n+\frac{bdt}{2m}(\beta^{n+1}+\beta^n)
\label{rV_displacement}
\nonumber \\
\end{eqnarray}
with the associated velocity given by
\begin{eqnarray}
v^n & = &
\frac{a}{b^2}\frac{r^{n+1}-r^{n-1}}{2dt}-\frac{\alpha}{2m}\frac{bdt^3}{m}f^n
\nonumber \\ &&
+\frac{bdt}{2m}(a\beta^{n+1}-\beta^n) \, . \label{rV_velocity}
\end{eqnarray}

Eq.~(\ref{rV_displacement}) shows that our formulation is also in
close proximity to the BBK scheme \cite{Brunger_1984}, which is of a
position-Verlet-type. (Notice that it is essential to have the proper
form for the accompanying velocity in order to determine if a
position formulation is the same as a velocity explicit form.)
However, unlike our Eq.~(\ref{rV_displacement}), the BBK scheme
employs only a single stochastic number for the two time steps covered
in the displacement equation. In fact, the methods of Melchionna and
BBK, while not identical, are closely related, since the use of a single
stochastic number
in the position representation (BBK) translates into applying two
random numbers in the velocity equation
(Melchionna). Conversely, in our scheme, we have a single random
number in the velocity formulation, and two in the
position representation covering two time steps.

\section{Linear Analysis}
In order to evaluate the general applicability of the above method, we
calculate key statistical measures of the numerical scheme for some
characteristic linear cases \cite{Mishra_1996}, and compare
them to known statistical-mechanical behavior of the true Langevin system.

\subsection{Thermal Diffusion, $f=0$}
The first basic property to investigate is the diffusive behavior of a
particle moving in a flat potential ($f=0$) at temperature $T$. In
this case, Eq.~(\ref{rVv_vs2}) reads:
\begin{eqnarray}
v^{n+1} & = & av^n+\frac{b}{m}\beta^{n+1} \; .  \label{rVv_flat_0}
\end{eqnarray}
By using the same equation to express $v^n$ in terms of $v^{n-1}$ and
$\beta^n$, and by repeating this procedure until reaching the initial
velocity $v^0$, one arrives at the relation
\begin{eqnarray}
v^{n} & = &
a^nv^0+\frac{b}{m}\sum_{k=0}^{n-1}a^k\beta^{n-k}\label{rVv_flat_2}\\ &
= & a^nv^0+\frac{b\sqrt{2\alpha
k_BTdt}}{m}\sum_{k=0}^{n-1}a^k\sigma^{n-k} \label{rVv_flat_3}
\end{eqnarray}
where $\sigma$ is a standard Gaussian random number with
$\langle\sigma\rangle=0$ and $\langle\sigma^2\rangle=1$, the
superscript denoting different realizations of this variable such that
$\langle\sigma^n\sigma^l\rangle=\delta_{nl}$.  Summing over the random
numbers in Eq.~(\ref{rVv_flat_3}) yields another Gaussian random
number,  and in combination with Eq.~(\ref{b_definition}),  one arrives at
\begin{eqnarray}
v^n & = & a^nv^0+\sqrt{1-a^{2n}}{\sqrt{\frac{k_BT}{m}}}\;\sigma \; .
\end{eqnarray}
For large $n$, beyond the transients from the initial condition
($a^n\ll1$), we find that the velocity is characterized by a Gaussian
(Maxwell-Boltzmann) distribution with zero mean and
\begin{eqnarray}
\langle(v^n)^2\rangle & = & \frac{k_BT}{m} , \label{rVv_vs2_therm}
\end{eqnarray}
which results in reproducing the exact expectation for the average
kinetic energy (thermal energy)
\begin{eqnarray}
E_k & = & \frac{1}{2}m\langle(v^n)^2\rangle \; = \; \frac{1}{2}k_BT.\label{flat_kin}
\end{eqnarray}
Complementing this result, we turn to the displacement coordinate
Eq.~(\ref{rVp_vs2}) for $f=0$:
\begin{eqnarray}
r^n & = & r^{n-1}+bdtv^{n-1}+\frac{bdt}{2m}\beta^n \; . \label{rVp_flat_0}
\end{eqnarray}
Inserting Eq.~(\ref{rVv_flat_3}) for $v^{n-1}$, and repeating the
procedure until reaching the initial position $r^0$, we obtain the result
\begin{eqnarray}
r^n & = & r^0+\frac{m}{\alpha}\left(1-a^n\right)v^0
 \nonumber \\ &+&
\frac{b}{\alpha}\left[\frac{\alpha
dt}{2m}\beta^n+\sum_{k=1}^{n-1}\left(\frac{1}{b}-a^k\right)\beta^{n-k}\right]. \label{rVp_flat_1}
\end{eqnarray}
By using the defintion of $a$ (\ref{a_definition}) and $b$
(\ref{b_definition}), we find that for large $n$ (such that $a^n\ll1$),
Eq.~(\ref{rVp_flat_1}) can be written as
\begin{eqnarray}
r^n & = & r^0+\frac{m}{\alpha}v^0+\sqrt{ndt\frac{2k_BT}{\alpha}}\;\sigma. 
\label{r_large_n}
\end{eqnarray}
The second term in Eq.~(\ref{r_large_n}), which is the transient
ballistic displacement, matches exactly the corresponding value
predicted by the Langevin solution [second term on the r.h.s of
Eq.~(\ref{analytic_r}) for $dt\rightarrow \infty$]. More importantly,
we find that our scheme results in a simulated diffusion coefficient
\begin{eqnarray}
D & = & \lim_{ndt\rightarrow\infty}\frac{\langle r^n-r^0\rangle^2}{2ndt} \; = \; \frac{k_BT}{\alpha}, \label{rVp_diffusion}
\end{eqnarray}
which agrees with Einstein's fluctuation-dissipation relationship for LD \cite{Parisi_1988}. Notice
that the exact results for fluctuations and diffusion obtained here
are independent of the magnitudes of the time step $dt$, damping $\alpha>0$, and temperature $T$.

\subsection{Thermal Harmonic Oscillator, $f=-\kappa r$}

Encouraged by the performance of the method for $f=0$, we now turn to
analyzing the method applied to a damped thermal harmonic oscillator,
where $f=-\kappa r$ represents a linear Hooke's spring with spring
constant $\kappa>0$. Our Eqs.~(\ref{rVp_vs2}) and (\ref{rVv_vs2}) now read
\begin{eqnarray}
\left(\begin{array}{c}r^{n+1}\\v^{n+1}\end{array}\right) & = & {\cal
V}\left(\begin{array}{c}r^{n}\\v^{n}\end{array}\right)+{\cal
N}\beta^{n+1}\label{HO_map0}
\end{eqnarray}
where
\begin{eqnarray}
{\cal V} & = & \left(\begin{array}{cc} 1-b\frac{\Omega_0^2dt^2}{2} &
bdt \\ -b\Omega_0^2 dt (1-\frac{\Omega_0^2dt^2}{4}) &
a-b\frac{\Omega_0^2dt^2}{2}\end{array}\right)\label{HO_Verlet_0}\\
{\cal N} & = &
\frac{b}{m}\left(\begin{array}{c}\frac{dt}{2}\\1-\frac{\Omega_0^2dt^2}{4}\end{array}\right)\label{HO_Noise_0},
\end{eqnarray}
and where $\Omega_0=\sqrt{\kappa/m}$ is the resonance frequency of the
undamped continuous-time oscillator. The matrix ${\cal V}$ has the
eigenvalues $\Lambda_\pm$:
\begin{eqnarray}
\Lambda_\pm & = &
b\left(1-\frac{\Omega_0^2dt^2}{2}\right)\pm\sqrt{b^2\left(1-\frac{\Omega_0^2dt^2}{2}\right)^2-a},
\nonumber \\
\label{eigen_0}
\end{eqnarray}
from where we can evaluate the formal
stability limit $\Omega_0dt<2$, consistent with the requirement $|\Lambda_\pm|<1$. Notice that $\Lambda_+\Lambda_-=a$ for all parameter values.

We analyze the basic thermodynamic properties of the thermal harmonic oscillator by perpetuating the map Eq.~(\ref{HO_map0})
from initial conditions $r^0$ and $v^0$, 
\begin{eqnarray}
\left(\begin{array}{c}r^{n}\\v^{n}\end{array}\right) & = & {\cal
V}\left(\begin{array}{c}r^{n-1}\\v^{n-1}\end{array}\right)+{\cal
N}\beta^{n}\label{HO_map1}\\ & = & {\cal
V}^n\left(\begin{array}{c}r^{0}\\v^{0}\end{array}\right)+\sum_{k=0}^{n-1}{\cal
V}^k{\cal N}\beta^{n-k}\label{HO_map2}
\end{eqnarray}
where the $k$th iteration is conveniently given by the unitary transformation
\begin{eqnarray}
{\cal V}^k & = & {\cal U}\left(\begin{array}{cc}\Lambda_+^k & 0 \\ 0 &
\Lambda_-^k\end{array}\right){\cal U}^{-1}. \label{HO_trans_k}
\end{eqnarray}

\begin{figure}
[ptb]
\begin{center}
\includegraphics[trim=60 90 90 30,
natheight=8.5in,
natwidth=11.0in,
height=3.0in,
]%
{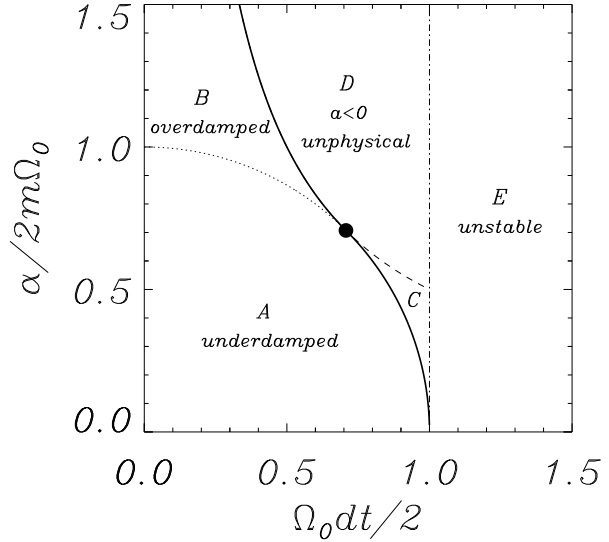}%
\caption{Sketch of the characteristic regimes for the Verlet integrator applied to the damped harmonic oscillator. Regimes are defined by the eigenvalues given in Eq.~(\ref{eigen_0}). Solid curve is the boundary between physical (left) and unphysical (right) behavior. Vertical dash-dotted line is the formal stability boundary of the Verlet method, $\Lambda_-=-1$. Marker~$\bullet$ indicates where $\Lambda_\pm=0$.
Regimes are: (A) Underdamped, $\Lambda_\pm$ complex; (B) Overdamped, $0<a<\Lambda_\pm<1$ real; (C)  $-1<\Lambda_\pm<-a<0$ real; (D) $\Lambda_\pm$ real, $\Lambda_-$ negative, $\Lambda_+$ positive ($a<0$); (E) Unstable, $\Lambda_-<-1$.}%
\label{Fig1}%
\end{center}
\end{figure}

The two characteristic parameters of the Verlet method applied to a damped harmonic oscillator are
$\alpha/2m\Omega_0$ and $\Omega_0dt$. Figure \ref{Fig1} displays the five different regimes of simulated
oscillator behavior as a function of those two parameters.
The two physically relevant dynamical regimes, underdamped (A) and overdamped (B), are considered first.

\subsubsection{Regime A, Underdamped Dynamics}

The oscillator is always stable in this regime, where
$(\alpha/2m)^2<\Omega_0^2(1-\Omega_0^2dt^2/4)$, and the eigenvalues can be
conveniently written
\begin{eqnarray}
\Lambda_\pm & = & \sqrt{a}\,e^{\pm i\Omega_Vdt}.
\end{eqnarray}
Here, $\Omega_V$ is the underdamped Verlet-oscillator resonance frequency given by
\begin{eqnarray}
\sqrt{a}\cos{\Omega_Vdt} & = &
b\left(1-\frac{\Omega_0^2dt^2}{2}\right)\label{eigen_cos_ud}\\
\sqrt{a}\sin{\Omega_Vdt} & = &
bdt\sqrt{\Omega_0^2\left(1-\frac{\Omega_0^2dt^2}{4}\right)-\left(\frac{\alpha}{2m}\right)^2}
\, ,
\nonumber \\ 
\label{eigen_sin_ud}
\end{eqnarray}
such that
\begin{widetext}
\begin{eqnarray}
{\cal V} & = & \left(\begin{array}{cc} \sqrt{a}\cos{\Omega_Vdt}+1-b & bdt\\ -bdt[\frac{a}{b^2dt^2}\sin^2{\Omega_Vdt}+(\frac{\alpha}{2m})^2] & \sqrt{a}\cos{\Omega_Vdt}+a-b \end{array}\right).
\end{eqnarray}
The diagonalizing transformation is then given by
\begin{eqnarray}
{\cal U} & = & \frac{1}{|u|}\left(\begin{array}{cc}1 & 1 \\
-\frac{\alpha}{2m}+i\frac{\sqrt{a}}{bdt}\sin\Omega_Vdt &
-\frac{\alpha}{2m}-i\frac{\sqrt{a}}{bdt}\sin\Omega_Vdt\end{array}\right)\label{HO_trans_U_ud}
\end{eqnarray}
with $|u|$ being the normalization of the eigenvectors of ${\cal V}$ that appear as the column vectors in ${\cal U}$. The $k$th
time step can then be expressed from Eq.~(\ref{HO_trans_k}) in the explicit form:
\begin{eqnarray}
{\cal V}^k & = & \frac{bdt}{\sqrt{a}\sin{\Omega_Vdt}}\sqrt{a}^k
\left(\begin{array}{cc}
\frac{\sqrt{a}}{bdt}\sin{\Omega_Vdt}\cos{\Omega_Vkdt}+\frac{\alpha}{2m}\sin{\Omega_Vkdt}
& \sin{\Omega_Vkdt} \\
-(\frac{a}{b^2dt^2}\sin^2\Omega_Vdt+(\frac{\alpha}{2m})^2)\sin{\Omega_Vkdt}
&
\frac{\sqrt{a}}{bdt}\sin{\Omega_Vdt}\cos{\Omega_Vkdt}-\frac{\alpha}{2m}\sin{\Omega_Vkdt}
\end{array}\right) , \label{Eq:eq_Vn}
\end{eqnarray}
from where we obtain the explicit form of the expression of interest in Eq.~(\ref{HO_map2})
\begin{eqnarray}
{\cal V}^k{\cal N}\beta^{n-k} & = & \frac{bdt\sqrt{2\alpha k_BT}}{2m\sqrt{a}\sin\Omega_Vdt}\sqrt{a}^k
\left(\begin{array}{c}
\sqrt{a}\sin{\Omega_Vdt}\cos{\Omega_Vkdt}+(1+\sqrt{a}\cos{\Omega_Vdt})\sin{\Omega_Vkdt}
\\
(1-\frac{\Omega_0^2dt^2}{4})\frac{2}{dt}[\sqrt{a}\sin{\Omega_Vdt}\cos{\Omega_Vkdt}-(1-\sqrt{a}\cos{\Omega_Vdt})\sin{\Omega_Vkdt}]
\end{array}\right)\sigma^{n-k} \, . \nonumber 
\\ 
\label{Eq:eq_HNoise}
\end{eqnarray}
\end{widetext}
The thermodynamic limit of
$\langle(r^n)^2\rangle$ and $\langle(v^n)^2\rangle$ can now be found by inserting Eq.~(\ref{Eq:eq_HNoise}) into
Eq.~(\ref{HO_map2}), whereafter the square sum can be evaluated since each term in the summation is an independent Gaussian random number. Taking the limit $n\rightarrow\infty$ this yields for the variance of the displacement:
\begin{widetext}
\begin{eqnarray}
\langle(r^n)^2\rangle & = & \frac{\alpha k_BTb^2dt^3}{2m^2a\sin^2{\Omega_Vdt}} \; \Big[\frac{1}{2}(1+a+2\sqrt{a}\cos{\Omega_Vdt})  \sum_{k=0}^\infty a^k \nonumber \\
&& - \frac{1}{2}(1+a\cos{2\Omega_Vdt}+2\sqrt{a}\cos{\Omega_Vdt})\sum_{k=0}^\infty a^k \cos{2\Omega_Vkdt}\nonumber \\
&&+ \sqrt{a}\sin{\Omega_Vdt}(1+2\sqrt{a}\cos{\Omega_Vdt})\sum_{k=0}^\infty a^k \sin{2\Omega_Vkdt} \Big] \nonumber \\
& = & \frac{\alpha k_BTb^2dt^3}{2m^2a\sin^2{\Omega_Vdt}} \; \Big[\frac{1}{2}\frac{4b(1-\frac{\Omega_0^2dt^2}{4})}{1-a}
-\frac{(1-a)\sqrt{a}\cos{\Omega_Vdt}+\frac{1}{2}(1-a^2)}{1+a^2-2a\cos{2\Omega_Vdt}}\Big] \nonumber \\
&=& \; \frac{\alpha k_BTb^2dt^3}{2m^2a\sin^2{\Omega_Vdt}} \; \frac{\Omega_0^2(1-\frac{\Omega_0^2dt^2}{4})-(\frac{\alpha}{2m})^2}{\frac{\alpha dt}{2m}\Omega_0^2} \nonumber \\
&=& \; \frac{k_BT}{m\Omega_0^2}.\label{HO_var_r}
\end{eqnarray}
We similarly obtain the result for the statistical variance of the velocity:
\begin{eqnarray}
\langle(v^n)^2\rangle & = & \frac{2\alpha k_BTb^2dt}{m^2a\sin^2{\Omega_Vdt}}\left(1-\frac{\Omega_0^2dt^2}{4}\right)^2
 \Big[\frac{1}{2}(1+a-2\sqrt{a}\cos{\Omega_Vdt})\sum_{k=0}^\infty a^k\nonumber \\
&&-\frac{1}{2}(1+a\cos{2\Omega_Vdt}-2\sqrt{a}\cos{\Omega_Vdt})\sum_{k=0}^\infty a^k\cos{2\Omega_Vkdt}\nonumber \\
&&+\sqrt{a}\sin{\Omega_Vdt}(1-\sqrt{a}\cos{\Omega_Vdt})\sum_{k=0}^\infty a^k\sin{2\Omega_Vkdt}\Big] \nonumber \\
&=&  \frac{2\alpha k_BTb^2dt}{m^2a\sin^2{\Omega_Vdt}}\left(1-\frac{\Omega_0^2dt^2}{4}\right)^2 
\Big[\frac{b\frac{\Omega_0^2dt^2}{2}}{1-a}+\frac{(1-a)\sqrt{a}\cos{\Omega_Vdt}-\frac{1}{2}(1-a^2)}{1+a^2-2a\cos{2\Omega_Vdt}}\Big] \nonumber \\
&=&  \frac{2\alpha k_BTb^2dt^2}{m^2a\sin^2{\Omega_Vdt}}\left(1-\frac{\Omega_0^2dt^2}{4}\right) 
\frac{\Omega_0^2(1-\frac{\Omega_0^2dt^2}{4})-(\frac{\alpha}{2m})^2}{4\frac{\alpha}{2m}} \nonumber \\
& = & \frac{k_BT}{m}\left(1-\frac{\Omega_0^2dt^2}{4}\right). \label{HO_var_v}
\end{eqnarray}
\end{widetext}

The two variances are noteworthy, since evaluating the average potential and kinetic energies gives
\begin{eqnarray}
E_p & = & \frac{1}{2}\kappa\langle(r^n)^2\rangle \; = \;
\frac{1}{2}k_BT \label{HO_res_Ep_ud}\\ E_k & = &
\frac{1}{2}m\langle(v^n)^2\rangle \; = \;
\frac{1}{2}k_BT\left(1-\frac{\Omega_0^2dt^2}{4}\right) \, .
\label{HO_res_Ek_ud}
\end{eqnarray}
These hold true for any parameter choice in the underdamped regime.
Notably, Eq.~(\ref{HO_res_Ep_ud}) implies that our method
produces the {\it exact} statistical distribution regardless of time
step, frequency (potential curvature), damping parameter, or
temperature. Therefore, it is reasonable to expect that this method will
provide correct Boltzmann distribution in thermodynamic
equilibrium when simulating complex many-particle systems that may have a
multiple of participating frequencies. To our knowledge, this
important feature has not previously been reported for a numerical integrator
of the Langevin equation \cite{Skeel_2003}.

Complementing the Boltzmann distribution of the displacement is the
variance of the velocity, Eq.~(\ref{HO_res_Ek_ud}). Despite the obvious
discrepancy between the true kinetic energy of the Langevin equation
and the one shown for our
algorithm, we submit that the presented result is the best possible
for a method that builds on the discretized Verlet formalism. The reduction
in the variance of the velocity by a
factor of $(1-\Omega_0^2dt^2/4)$ does not arise from the treatment of
friction and noise, as implied by the fact that this factor depends on neither damping nor temperature.
Instead, the observed deviation arises from the approximation of the
potential curvature introduced by the discrete integrator. This is consistent with the
well-known inherent discrepancy between displacement and
associated velocity that causes the periodic deviations with magnitude
$dt^2$ from strict energy conservation in a simulated harmonic
oscillator. This can be explicitly demonstrated by using the Verlet equations
(\ref{vverlet_pos}) and (\ref{vverlet_vel}) for an undamped harmonic
oscillator with initial conditions $r^0$ and $v^0$. The result is
\begin{eqnarray}
&&\left(\begin{array}{c}r^n \\ v^n\end{array}\right) \; = \;
\left(\begin{array}{c}r^0\\v^0\end{array}\right)\cos\Omega_Vndt 
\nonumber  \\ &&
+ \frac{1}{\sqrt{\Omega_0^2\left(1-\frac{\Omega_0^2dt^2}{4}\right)}}\left(\begin{array}{c}v_0\\-\Omega_0^2(1-\frac{\Omega_0^2dt^2}{4})r_0\end{array}\right)\sin\Omega_Vndt
\nonumber \\ 
\label{verlet_mismatch}
\end{eqnarray}
which illustrates that the Verlet velocity (momentum) is depressed and is not exactly the
conjugated coordinate to the displacement, and that the discrepancy is
related to
the proportionality seen between the two thermodynamic expressions
(\ref{HO_res_Ep_ud}) and (\ref{HO_res_Ek_ud}).\\

\subsubsection{Regime B, Overdamped Dynamics}

This regime is defined by the requirements
$(\alpha/2m)^2>\Omega_0^2(1-\Omega_0^2dt^2/4)$ and $\alpha dt/2m<1$ ( for $a>0$). The latter condition
is imposed to ensure $\Lambda_\pm>0$, which is necessary for physically meaningful dynamics, where only monotonic decay
is possible in the overdamped regime for $T=0$.

The matrix ${\cal V}$ appearing in Eq.~(\ref{HO_map0}), which is now expressed by
\begin{widetext}
\begin{eqnarray}
{\cal V} & = & \left(\begin{array}{cc} \sqrt{a}\cosh{\lambda_Vdt}+1-b & bdt\\ -bdt[(\frac{\alpha}{2m})^2-\frac{a}{b^2dt^2}\sinh^2{\lambda_Vdt}] & \sqrt{a}\cosh{\lambda_Vdt}+a-b \end{array}\right),
\end{eqnarray}
has the eigenvalues
\begin{eqnarray}
\Lambda_\pm & = & \sqrt{a}\,e^{\pm \lambda_Vdt},\label{Lambda_od}
\end{eqnarray}
where $\lambda_V$ is given by
\begin{eqnarray}
\sqrt{a}\cosh{\lambda_Vdt} & = &
b\left(1-\frac{\Omega_0^2dt^2}{2}\right)\label{eigen_cosh}\\
\sqrt{a}\sinh{\lambda_Vdt} & = &
bdt\sqrt{\left(\frac{\alpha}{2m}\right)^2-\Omega_0^2\left(1-\frac{\Omega_0^2dt^2}{4}\right)}
\, . \label{eigen_sinh}
\end{eqnarray}
The matrix ${\cal V}$  can be diagonalized (see Eq.~(\ref{HO_map0})) using the transformation matrix
\begin{eqnarray}
{\cal U} & = & \left(\begin{array}{cc}\frac{1}{|u_+|} & \frac{1}{|u_-|} \\
\frac{1}{|u_+|}[-\frac{\alpha}{2m}+\frac{\sqrt{a}}{bdt}\sinh\lambda_Vdt] &
\frac{1}{|u_-|}[-\frac{\alpha}{2m}-\frac{\sqrt{a}}{bdt}\sinh\lambda_Vdt]\end{array}\right)
\label{HO_trans_U_od}
\end{eqnarray}
with $|u_+|$ and $|u_-|$ being the normalizations of the two real eigenvectors of ${\cal V}$. This leads to the explicit results:
\begin{eqnarray}
&&{\cal V}^k \; = \;
\frac{bdt}{\sqrt{a}\sinh{\lambda_Vdt}}\sqrt{a}^k
\left(\begin{array}{cc}
\frac{\sqrt{a}}{bdt}\sinh{\lambda_Vdt}\cosh{\lambda_Vkdt}+\frac{\alpha}{2m}\sinh{\lambda_Vkdt}
& \sinh{\lambda_Vkdt} \\
-((\frac{a}{2m})^2-\frac{a}{b^2dt^2}\sinh^2\lambda_Vdt)\sinh{\lambda_Vkdt}
&
\frac{\sqrt{a}}{bdt}\sinh{\lambda_Vdt}\cosh{\lambda_Vkdt}-\frac{\alpha}{2m}\sinh{\lambda_Vkdt}
\end{array}\right) \nonumber 
\\ 
\label{Eq:eq_Vn_od}
\end{eqnarray}
and
\begin{eqnarray}
&&{\cal V}^k{\cal N}\beta^{n-k} \; = \; \frac{bdt\sqrt{2\alpha k_BT}}{2m\sqrt{a}\sinh\lambda_Vdt}\sqrt{a}^k
\; \times\nonumber \\ && 
\left(\begin{array}{c}
\sqrt{a}\sinh{\lambda_Vdt}\cosh{\lambda_Vkdt}+(1+\sqrt{a}\cosh{\lambda_Vdt})\sinh{\lambda_Vkdt}
\\
(1-\frac{\Omega_0^2dt^2}{4})\frac{2}{dt}[\sqrt{a}\sinh{\lambda_Vdt}\cosh{\lambda_Vkdt}-(1-\sqrt{a}\cosh{\lambda_Vdt})\sinh{\lambda_Vkdt}]
\end{array}\right)\sigma^{n-k} \, .  
\label{Eq:eq_HNoise_od}
\end{eqnarray}
\end{widetext}
Following the same procedure used in the underdamped case above, we insert Eq.~(\ref{Eq:eq_HNoise_od}) into 
Eq.~(\ref{HO_map2}), and evaluate the variances of the displacement and velocity in the thermodynamic limit for $n\rightarrow\infty$
by calculating the square sum of the amplitudes of the independent stochastic numbers of each term. After some calculations, we arrive at
\begin{widetext}
\begin{eqnarray}
\langle(r^n)^2\rangle & = & \frac{\alpha k_BTb^2dt^3}{2m^2a\sinh^2{\lambda_Vdt}} \; \Big[-\frac{1}{2}(1+a+2\sqrt{a}\cosh{\lambda_Vdt})  \sum_{k=0}^\infty a^k \nonumber \\
&&+\frac{1}{2}(1+a\cosh{2\lambda_Vdt}+2\sqrt{a}\cosh{\lambda_Vdt})\sum_{k=0}^\infty a^k \cosh{2\lambda_Vkdt}\nonumber \\
&&+\sqrt{a}\sinh{\lambda_Vdt}(1+2\sqrt{a}\cosh{\lambda_Vdt})\sum_{k=0}^\infty a^k \sinh{2\lambda_Vkdt} \Big] \nonumber \\
& = & \frac{\alpha k_BTb^2dt^3}{2m^2a\sinh^2{\lambda_Vdt}} \; \Big[-\frac{1}{2}\frac{4b(1-\frac{\Omega_0^2dt^2}{4})}{1-a}
+\frac{(1-a)\sqrt{a}\cosh{\lambda_Vdt}+\frac{1}{2}(1-a^2)}{1+a^2-2a\cosh{2\lambda_Vdt}}\Big] \nonumber \\
&=& \; \frac{\alpha k_BTb^2dt^2}{2m^2a\sinh^2{\lambda_Vdt}} \; \frac{(\frac{\alpha}{2m})^2-\Omega_0^2(1-\frac{\Omega_0^2dt^2}{4})}{\frac{\alpha}{2m}\Omega_0^2} \nonumber \\
&=& \; \frac{k_BT}{m\Omega_0^2},\label{HO_var_r_od}
\end{eqnarray}
which is exactly the same result as for the underdamped case Eq.~(\ref{HO_var_r}). We similarly obtain the result for the variance of the velocity:
\begin{eqnarray}
\langle(v^n)^2\rangle & = & \frac{2\alpha k_BTb^2dt}{m^2a\sinh^2{\lambda_Vdt}}\left(1-\frac{\Omega_0^2dt^2}{4}\right)^2
 \Big[-\frac{1}{2}(1+a-2\sqrt{a}\cosh{\lambda_Vdt})\sum_{k=0}^\infty a^k\nonumber \\
&&+\frac{1}{2}(1+a\cosh{2\lambda_Vdt}-2\sqrt{a}\cosh{\lambda_Vdt})\sum_{k=0}^\infty a^k\cosh{2\lambda_Vkdt}\nonumber \\
&&-\sqrt{a}\sinh{\lambda_Vdt}(1-\sqrt{a}\cosh{\lambda_Vdt})\sum_{k=0}^\infty a^k\sinh{2\lambda_Vkdt}\Big] \nonumber \\
&=&  \frac{2\alpha k_BTb^2dt}{m^2a\sinh^2{\lambda_Vdt}}\left(1-\frac{\Omega_0^2dt^2}{4}\right)^2 
\Big[\frac{b\frac{\Omega_0^2dt^2}{2}}{1-a}+\frac{(1-a)\sqrt{a}\cosh{\lambda_Vdt}-\frac{1}{2}(1-a^2)}{1+a^2-2a\cosh{2\lambda_Vdt}}\Big] \nonumber \\
&=&  \frac{2\alpha k_BTb^2dt^2}{m^2a\sinh^2{\lambda_Vdt}}\left(1-\frac{\Omega_0^2dt^2}{4}\right) 
\frac{\Omega_0^2(1-\frac{\Omega_0^2dt^2}{4})-(\frac{\alpha}{2m})^2}{4\frac{\alpha}{2m}} \nonumber \\
& = & \frac{k_BT}{m}\left(1-\frac{\Omega_0^2dt^2}{4}\right), \label{HO_var_v_od}
\end{eqnarray}
\end{widetext}
which is also identical to the comparable underdamped result from Eq.~(\ref{HO_var_v}). The averages of potential and kinetic energies
are therefore given by Eqs.~(\ref{HO_res_Ep_ud}) and (\ref{HO_res_Ek_ud}) also for the overdamped regime.
Thus, we can now conclude that our method
produces the {\it exact} Boltzmann distribution for any physical dynamics, underdamped or overdamped,
regardless of time
step, frequency (potential curvature), damping parameter, or
temperature.

\subsubsection{Regime C, $\Omega_V=\pi/dt$}

This somewhat unphysical regime, is characterized by conditions similar to the overdamped regime B; i.e., $\alpha dt/2m<1$ (for $a>0$) and $(\frac{\alpha}{2m})^2>\Omega_0^2(1-\frac{\Omega_0^2dt^2}{4})$. However, while regime B corresponds to $\Omega_0dt<\sqrt{2}$, regime C is defined for $\Omega_0dt>\sqrt{2}$, resulting in $\Lambda_\pm<0$. Thus, this regime
is typically reached for large time steps when simulating lightly damped dynamics, and
it is the pre-curser for violating formal stability, given by $\Omega_0dt<2$ (see Fig.~\ref{Fig1}).
At $T=0$ and with initial conditions $r^0\neq0$, the dynamics is characterized by $r^n$ and
$v^n$ alternating their signs every time step (i.e., $\Omega_V=\pi/dt$) with an exponentially decaying envelope.
For $T>0$, the thermodynamic properties of this regime can, therefore, be evaluated similarly
to that of the overdamped regime, since we here can write
\begin{eqnarray}
{\cal V}^k & = & (-1)^k\, {\cal U}\left(\begin{array}{cc}\Lambda_+^k & 0 \\ 0 & \Lambda_-^k\end{array}\right) {\cal U}^{-1}\, , \label{Regime_C}
\end{eqnarray}
where ${\cal U}$ and $\Lambda_\pm$ are given by Eqs.~(\ref{Lambda_od})-(\ref{HO_trans_U_od}). Apart from the alternating sign change in Eq.~(\ref{Regime_C}) this is identical to the corresponding mapping in regime B. This implies that Eqs.~(\ref{HO_var_r_od}) and (\ref{HO_var_v_od}) from regime B for the variances of $r^n$ and $v^n$ also apply here.
Thus,
we conclude that the thermodynamic results Eq.~(\ref{HO_res_Ep_ud}) and Eq.~(\ref{HO_res_Ek_ud}) are found
also in this regime, which is physically unreasonable, yet numerically accessible.

\subsubsection{Regime D, Unphysical}

This unphysical regime, $\alpha dt/2m>1$ (yet within the formal stability criterion $\Omega_0dt<2$),  is typically reached for large time steps when
simulating moderate to strongly damped
dynamics (see Fig.~\ref{Fig1}). Since this regime involves one
positive and one negative eigenvalue, we can map this onto the analysis from regime B with $\sqrt{a}\rightarrow\sqrt{|a|}$ and
$\lambda_V\rightarrow-\lambda_V$ to complete the analysis
and obtain the Boltzmann distribution.

\section{Discussion and Conclusion}

We have presented 
a new Verlet-type algorithm for simulating Langevin dynamics. The
method, written in the three different forms Eqs.~(\ref{rVp_vs1})-(\ref{rVv_vs1}), Eqs.~(\ref{rVp_vs2})-(\ref{rVv_vs2}) or Eqs.~(\ref{rV_displacement})-(\ref{rV_velocity}), is aligned with other published methods
that have demonstrated first, second, and third order accuracy for both simulated trajectories and derived thermodynamic quantities \cite{Brunger_1984,Skeel_2003,eb_1980,vgb_1982,Skeel_2002,rc_2003,Melchionna_2007,Bussi_2007,thalmann_2007,Mishra_1996}.
Linear analysis demonstrates that the
method of this paper is robust and capable of providing {\it exact} representation of both
diffusive behavior in a flat potential and the
Boltzmann distribution in a harmonic potential regardless of damping and frequency (curvature of the potential).
The derived exact distribution is obtained for any time step subject to the usual Verlet
stability limit and the condition $\alpha dt<2m$, which is a necessary requirement for
a meaningful attenuated trajectory. The method is very simple and in the
usual Verlet structure, which means that it can be readily implemented
for any Langevin application, including molecular dynamics of
many-particle systems with and without molecular constraints and other
commonly used modeling features. 

As a final note, we underline that a Verlet-simulated oscillator
may not measure the exact temperature from
the variance of the velocity (see Eq.~(\ref{HO_res_Ek_ud})). This deficiency, which is common to all Verlet-type methods,
arises from the known
discrepancy between displacement and momentum as conjugated variables. This means that using the
variance of the velocity for precisely assessing the temperature of a simulated system may
be counterproductive. The appropriate criterion for obtaining the desired temperature is the achievement of correct
statistical sampling, which is indeed given by our method.
 
\begin{acknowledgments}
NGJ acknowledges a very useful discussion with Zhaojun Bai. This project was supported in part by the US Department of Energy Project \# DE-NE0000536 000.
\end{acknowledgments}

\end{document}